\documentclass[sigconf,10pt]{acmart}

\usepackage{graphicx}
\usepackage{amsmath}
\usepackage{booktabs}
\renewcommand\footnotetextcopyrightpermission[1]{}
\usepackage{natbib}
\usepackage{algorithm}
\usepackage{algpseudocode}

\title{Environment-Aware Dynamic Pruning for Pipelined Edge Inference}
\author{Austin O'Quinn, Conor Snedeker, Siyuan Zhang, Jenna Kline}
\affiliation{%
  \institution{The Ohio State University, United States}
  \city{}
  \country{}
}
\email{{oquinn.18, snedeker.31, zhang.15333, kline.377}@osu.edu}

\acmConference[EdgeSys '25]{8th EdgeSys}{March 31, 2025}{Rotterdam, Netherlands}
\acmYear{2025}
\copyrightyear{2025}
\acmDOI{}
\acmISBN{}

\begin{document}
\pagestyle{plain}

\begin{abstract}
\emph{
IoT and edge-based inference systems require unique solutions to overcome resource limitations and unpredictable environments. In this paper, we propose an environment-aware dynamic pruning system that handles the unpredictability of edge inference pipelines. While traditional pruning approaches can reduce model footprint and compute requirements, they are often performed only once, offline, and are not designed to react to transient or post-deployment device conditions. Similarly, existing pipeline placement strategies may incur high overhead if reconfigured at runtime, limiting their responsiveness. Our approach, allows slices of a model—already placed on a distributed pipeline—to be ad-hoc pruned as a means of load-balancing. To support this capability, we introduce two key components: (1) novel training strategies that endow models with robustness to post-deployment pruning, and (2) an adaptive algorithm that determines the optimal pruning level for each node based on monitored bottlenecks. In real-world experiments, on a Raspberry Pi 4B cluster running camera-trap workloads, our method achieves a $1.5 \times$ speedup and a $3 \times$ improvement in service-level objective (SLO) attainment, all while maintaining high accuracy.
}
\end{abstract}
\maketitle
\section{Introduction}
\label{sec:intro}
AI is rapidly expanding into the edge, creating a heightened demand for high-performance, low-latency inference across distributed, resource-limited heterogeneous devices. In these edge environments, practitioners often turn to model parallelism: partitioning and distributing large models across multiple devices to fit memory constraints and to better handle bursty workloads. Recent strategies and systems such as DistrEdge \citep{hou2022distredge}, PipeEdge \citep{hu2022pipeedge}, and AlpaServe \citep{li2023alpa} demonstrate the potential of intelligent model placement. However, while these systems allocate model slices effectively, they are not designed to adapt rapidly and frequently to the dynamic conditions typical of edge environments.

\begin{figure}[htbp]
\centering
\includegraphics[width=\linewidth]{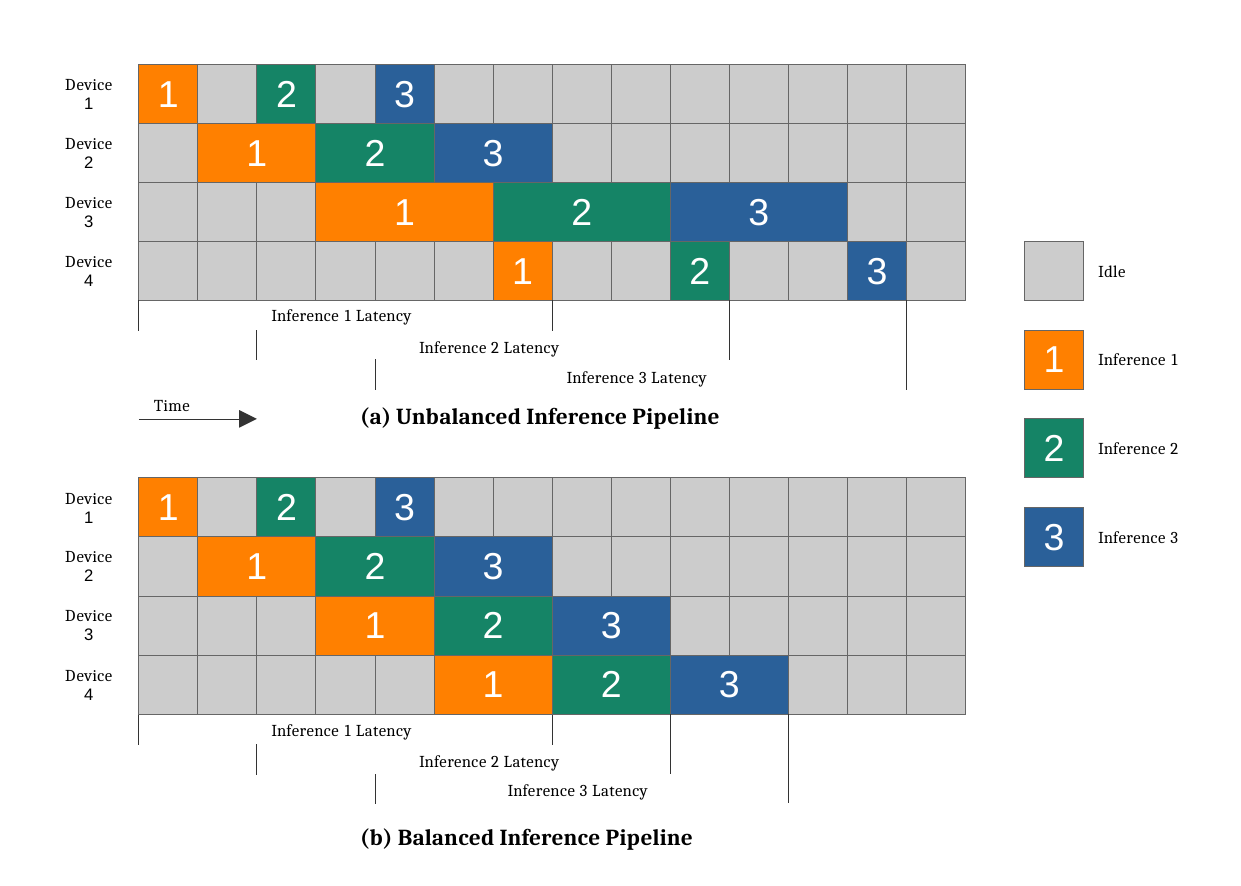}
\caption{Pipeline inference comparison with three stages across four edge devices. (a) An imbalanced pipeline, where stage 1 is notably slower, increasing latency for subsequent inferences, (b) a balanced pipeline with evenly distributed stage durations yielding improved latency and throughput.}
\label{fig:balance_pipeline}
\end{figure}

Despite intelligent model placement, distributed edge systems face distinct challenges that often lead to pipeline imbalance and performance degradation. Hardware heterogeneity makes uniform deployment strategies suboptimal, as layer-wise splitting can disproportionately tax slower devices. Additionally, transient conditions such as data uploads, background tasks, or dual-use devices cause unpredictable slowdowns at individual pipeline stages. Even with an initially well-optimized model placement, these episodic bottlenecks trigger load imbalances, resulting in some nodes being overloaded while others are near idle. Re-partitioning to correct this is costly and, in memory-limited devices, may be impractical.

While GPUs are the dominant computing platform in data centers, real-world edge and IoT deployments typically rely on CPU-based devices due to practical constraints. Battery-powered applications require low power consumption, and CPUs generally draw far less energy than GPU solutions. Moreover, the higher cost of GPU-enabled hardware is difficult to justify for deployments with large numbers of short-lifespan devices. CPU-based platforms such as the Raspberry Pi 4B offer a pragmatic balance of performance, cost-effectiveness, and power efficiency for far-edge inference.

To address these hardware constraints and the variable operating conditions of the edge, we propose environment-aware dynamic pruning, which adjusts model architecture in real-time based on device-level performance metrics. While traditional pruning has long been leveraged to compress models and accelerate devices prior to deployment, on-the-fly pruning remains rare—largely because accuracy can suffer when fine-tuning is infeasible. Our approach overcomes this drawback with two key innovations. First, we introduce training techniques that preserve a model’s accuracy on a continuum of pruning levels, eliminating the need for extensive re-training each time pruning is applied. Second, we design an online pruning algorithm that monitors metrics such as queuing delay and prunes under-performing pipeline stages without interrupting service. Our solution employs structured pruning with model surgery, removing entire channels or filters rather than merely masking them. Pruned portions can be dynamically reactivated when conditions improve, allowing the pruning level to be lowered, restoring full model capacity as resources become available. By excising these structures, we reduce the load on bottleneck devices and maintain balanced pipeline performance.

\paragraph{Roadmap.} Our paper is organized as follows:
Section~\ref{sec:design} details our system design and dynamic pruning mechanism and describes the implementation. 
Section~\ref{sec:evaluation} presents an experimental evaluation on a real-world workload. 
Section~\ref{sec:future} is a discussion of our orientation toward future work. 
Finally, Section~\ref{sec:conclusion} concludes the paper.
\section{System Design and Dynamic Pruning Mechanism}
\label{sec:design}

\begin{figure*}[htp]
  \centering
  \includegraphics[width=1\textwidth]{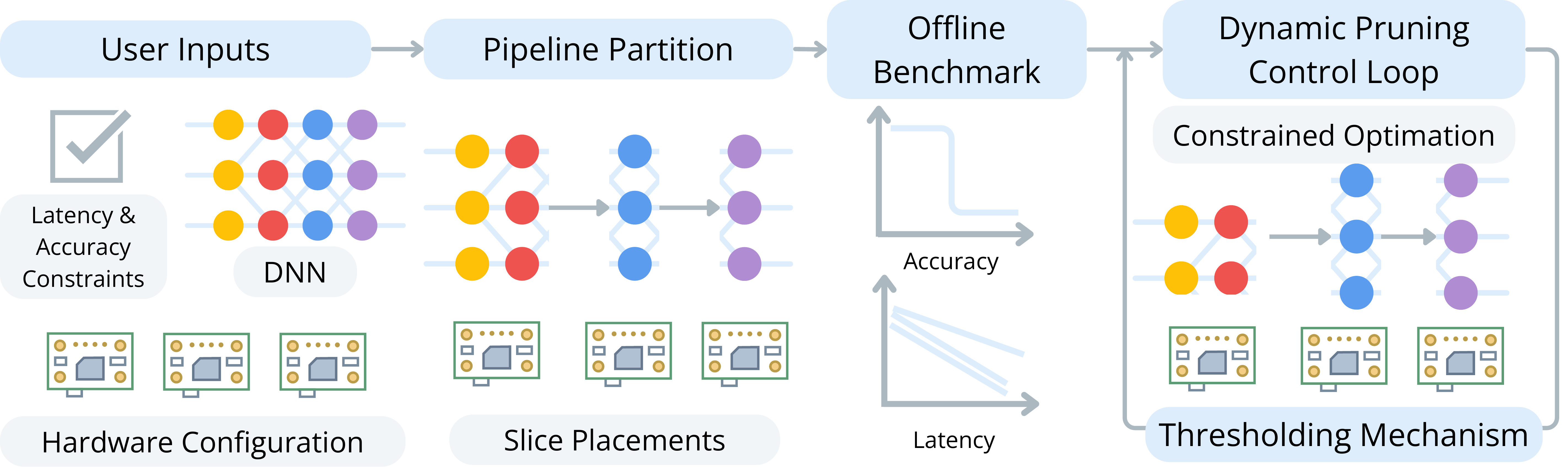}
  \caption{System design overview: user inputs, pipeline partition, offline benchmark, and dynamic pruning control loop.}
  \label{fig:system-overview}
\end{figure*}


Our end-to-end solution for distributed, pipelined inference on edge clusters is illustrated in Figure \ref{fig:system-overview}.  This approach combines an initial slice placement, offline benchmarking, a thresholding mechanism, and a constrained optimization step to determine pruning ratios. We implement \emph{slice placement} via a pipeline-parallel partitioning framework that handles heterogeneous devices.
 Next, a short \emph{offline benchmarking} phase fits slice-level latency and global accuracy curves at various pruning ratios. Our \emph{threshold  mechanism} decides when to prune, triggered by persistent end-to-end SLO violations. Next, we use \emph{constrained optimization} to decide how much to prune each slice, ensuring accuracy remains above a user-defined minimum. We use {Torch-Pruning} \cite{TorchPruning} to remove channels and filters.
With these components, we achieve a low-overhead, environment-aware approach that can adapt to transient device slowdowns or post placement imbalance—all without retraining or offline reconfiguration.


\subsection{Deployment and Implementation}
\label{sec:pipelineplacement}


Our system designates a controller node to execute pipeline partitioning, conduct dynamic pruning, and monitor system-wide latency. The controller assigns each device a model slice, enabling peer-to-peer activation exchanges. Once partitioning is set, devices listen for activation tensors from the preceding node and transmit outputs to the next. They also receive live pruning signals from the controller for performance optimization.

We embed our program in the central controller on Ray Serve \cite{ray_usenix} to handle model partitioning in the following way: First, we measure forward-pass time and peak memory usage for each layer or block on each Pi 4B. This only needs to be done once per model. Our system's dynamic programming routine then finds a slicing strategy that minimizes the pipeline's maximum stage latency via balancing heterogeneous devices. 

Once placement is finalized, each device loads its assigned slice — a contiguous set of layers ensuring computational balance. A full, unpruned copy of slice weights is stored locally for potential restoration. At runtime, device $i$ transmits data to device $i+1$, using pre-saved masks from slice $i$ to reconstruct pruned internal outputs. Pruning reduces output size, cutting communication time. By structuring slices contiguously, we minimize communication hops, making this approach well-suited for low-bandwidth edge networks. After deployment, dynamic pruning adjusts slices in real time to maintain efficiency.

\subsection{Short Benchmarking for Latency--Accuracy Curves}
\label{sec:benchmarking}
For each slice $i$ , as determined by our partitioning method, we prune it at a few discrete ratios 
and measure the resulting inference time $t_i(p_i)$. An example of these descrete ratios may be $p_i \in \{0,0.25,0.50,0.75,.90\}$. We then fit these points to the linear function
\[
t_i(p_i) \approx \alpha_i\,p_i + \beta_i
\]
which is used in our optimization routine. Although real hardware can exhibit mild non-linearity, we find a linear approximation is usually sufficient. Next, we prune the entire pipeline to various pruning ratios $\mathbf{p}=(p_1,\ldots,p_n)$ and measure the end-to-end accuracy. We then fit the results to a logistic curve $a(\mathbf{p})$ defined
\[
a(\mathbf{p}) 
= \frac{1}{1+\exp\Bigl(-\bigl(\sum_{i=1}^n \gamma_i\,p_i - \delta\bigr)\Bigr)}.
\]
These fitted curves and parameters $\{\alpha_i,\beta_i,\gamma_i,\delta\}$ for each slice $i$ are cached in our designated controller node, allowing for fast real-time pruning decisions without post-deployment measurements.

We rank the channels in a layer by $\ell_1$ norm importance and apply a uniform pruning ratio on each layer of a slice. This straight-forward pruning method allows our system to make fast pruning decisions from these fitted benchmark curves and reduces overhead. Furthermore, this manner of pruning makes it easier to predict how a sliced model will perform when each slice has distinct pruning ratios. We further expand on our methods, along with this latter point, in Section~\ref{sec:prunemethod}.
\begin{figure*}[htp]
\centering
\includegraphics[width=1\linewidth]{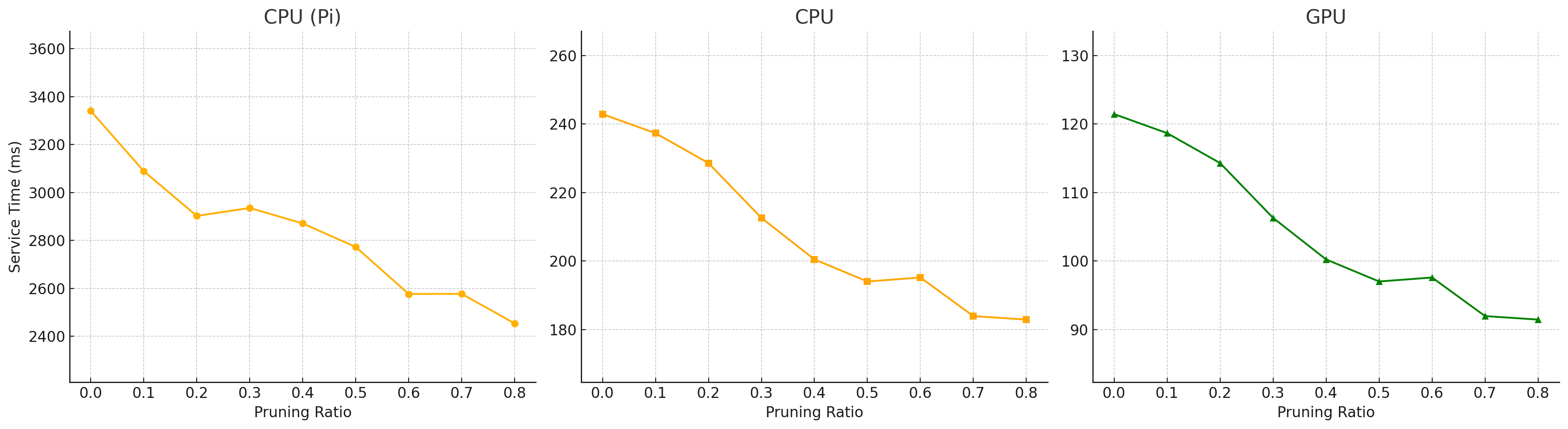}
\caption{Speedup curves for BioCLIP across different hardware platforms at various pruning ratios}\label{fig:cross-platform-speedup}
\end{figure*}
\subsection{Pruning Controller}
\label{sec:trigger}
During normal operation, the pipeline runs inference requests from slice $1$ to slice $n$. Then, we sample end-to-end latency to detect persistent slowdowns. We outline the threshold regime for this detection process below.

Note that the pruning ratio does \emph{not} equal the proportion of pruned parameters, but rather the proportion $r\in[0,1]$ of removed channels. That is, a pruning ratio of $0.5$ may lead to significantly more than half of all parameters in a model being pruned, depending on implementation. This behavior is expected as the number of pruned parameters when pruning channels will depend on the type of layers (e.g. convolutional, linear) being pruned and the individual inter-layer architecture of a model when using Torch-Pruning\footnote{For more information of Torch-Pruning  \url{https://github.com/VainF/Torch-Pruning/}.}.

\paragraph{Thresholds \emph{(}$\emph{\text{LAT}}_{\emph{\text{trigger}}}$, $\emph{\text{LAT}}_{\emph{\text{cooldown}}}$\emph{)}.}
If end-to-end latency exceeds $\text{LAT}_{\text{trigger}}$ for a short sustained window, typically seconds, we fire a pruning event. Usually, $\text{LAT}_{\text{trigger}}$ is set to the system SLO plus an overhead margin. After pruning, we will not attempt to prune again until the $\text{LAT}_{\text{cooldown}}$ 
timer expires. This hysteresis  prevents repeated pruning events for minor fluctuations and gives the system time to stabilize. Because CPU or GPU scheduling spikes may resolve quickly, we only prune under consistent overload. This ensures we preserve accuracy whenever possible and avoid overhead from frequent reconfiguration.

\paragraph{Selecting the Pruning Ratios.}
For simplicity we maintain six discrete pruning ratios per slice, ranging from $0$ (unpruned) to $1$ (fully pruned). We fit a linear slice-latency curve for each device as in Section~\ref{sec:benchmarking} along with a global logistic model for accuracy. Using these precomputed curves, the controller solves the optimization:
\[
\min_{\mathbf{p}} \sum_i (\alpha_i\,p_i + \beta_i) 
\quad \text{subject to } a(\mathbf{p}) \ge A_{\min},\ 
0 \le p_i \le 1
\]
where $p_i$ is the pruning ratio for slice $i$. Once an optimal $\mathbf{p}^*$ is found, the controller sends a ``prune now'' message to each device and the chosen ratio $p_i$ for the slice on device $i$. In our testing on the Pi~4B, this process typically took 25 ms to remove pruned channels on our model. Given that pruning events are rare, this overhead is negligible in comparison to the inference pipeline speedup it provides.

Since $t_i$ is approximately linear and $a(\mathbf{p})$ is logistic, we typically solve for $\mathbf{p}$ in one pass. First, we find $\mathbf{p}$ that satisfies $a(\mathbf{p}) = A_{\min}$, i.e. the largest possible pruning that does not breach the accuracy constraint. If that maximum is still insufficient to meet $\text{LAT}_{\text{cooldown}}$, the pipeline is infeasible for this hardware. If it overshoots, we do a small line search or ratio-based distribution, pruning more heavily on slices that yield the greatest latency reduction per unit accuracy cost $(\alpha_i/\gamma_i)$. Note that when the pipeline structure has multi-branch layers or more complex synergy among slices, a few gradient-descent steps easily find a feasible $\mathbf{p}^*$. Each iteration is fast, thanks to our pre-computed $\{\alpha_i,\beta_i,\gamma_i,\delta\}$.

\paragraph{Monitoring and Triggering SLO Checks.}
A key design point is how we detect violations of the end-to-end latency SLO. We timestamp inference requests at entry (the first slice) and exit (the final slice). The exit slice sends these timestamps back to the central controller. If the exit node notices that a significant fraction of requests exceed the user-defined SLO over a short window, typically measured in seconds, the central node initiates a more detailed resource probe. The controller aggregates these metrics, plus optional CPU utilization or queue length from each node, and decides whether we have a persistent overload rather than a brief spike. Once confirmed, it attempts to prune the pipeline. This mechanism avoids constant overhead from each device, instead focusing on end-to-end performance.

Overall, this design maintains high reliability, throughput, and accuracy under bursty or unpredictable pipeline loads. In Section~\ref{sec:evaluation} we present evaluations of our approach on real hardware testbeds (Raspberry Pis and RTX laptops), demonstrating that it achieves superior SLO adherence and speedups, \emph{without} retraining or offline reconfiguration.

\subsection{Pruning Method}\label{sec:prunemethod}
A primary goal of our training and pruning procedure is to deploy neural networks on the edge that can be heavily pruned with only minimal accuracy degradation. We must accomplish this with no fine-tuning steps after the network has been pruned. All on-device pruning is implemented with the existing features of Torch-Pruning.

\paragraph{Pruning-aware Training}
To conform with our goal of post-deployment dynamic pruning, we require a model to be pruned without subsequent fine-tuning. This cannot be accomplished with publicly available pretrained weights to the level of pruning required. To circumvent this issue we introduce training regimes that improve a model's robustness to pruning. In our preliminary experiments, we observe that smaller batch sizes, larger amounts of $\ell_2$-regularization, and training with more epochs all together instill this robustness in the studied models. 

\paragraph{Pruning Procedure.}
Our pruning methods are straightforward. 
This consists of taking the $\ell_1$ norm of all weights in each channel of a layer and ranking the channels in increasing order by norm size. The channels are pruned to a pruning ratio $r$ by removing the channels in the bottom $(100\cdot r)$\% of the importance ranking. The overhead for pruning a model in-place on an edge device with this method is minimal given the ease of computing the $\ell_1$ norm and storing an importance ranking over the channels of each layer.

\paragraph{Accuracy of Sliced Models.}\label{par:slicedmodelaccuracy}
The pruning decisions in one layer of a model slice are independent of the pruning decisions of adjacent layers. This layer-by-layer pruning decision streamlines the prediction of how the accuracy of a model will be impacted when separate slices are pruned to distinct pruning ratios. Our preliminary testing shows that a visual model sliced into $2$ or $3$ slices with varying pruning ratios typically has an accuracy no worse than a model with a uniform pruning ratio across all layers.

\section{Evaluation}
We verified our solution works on diverse hardware to obtain our results in Figure~\ref{fig:cross-platform-speedup}. However, we chose the Raspberry Pi 4B-8GB as our primary testing platform because this device is the one used in the deployment that processed our tested workload. We also believe the Pi 4B is a generally representative device for resource-constrained deployments such as camera traps \citep{zualkernan2022iot, jolles2022raspberrypi} and drones. 
For our end-to-end evaluation, we deployed a two-Raspberry Pi pipeline. Our evaluation emphasizes realism across all dimensions: we use production traces, representative edge hardware, the same CNN used to process our trace data, and SLOs derived from actual deployment requirements. 
\label{sec:evaluation}

\subsection{Accuracy Maintenance}

\begin{figure}[htp]
\centering
\includegraphics[width=1\linewidth]{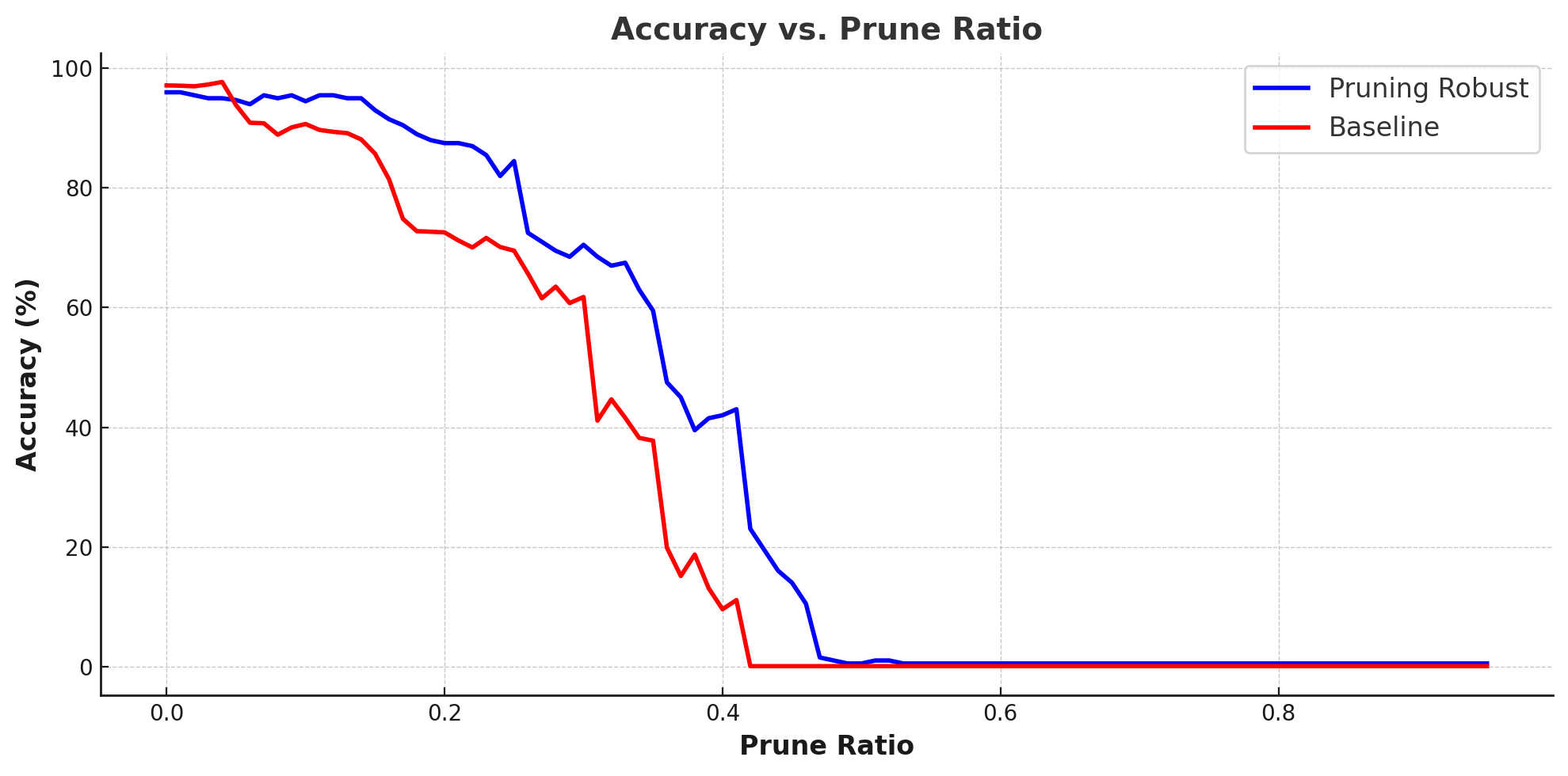}
\caption{Accuracy curves for BioCLIP tested on DSAIL Camera Trap data \cite{mugambi2022dsail} using two different sets of hyper-parameters }\label{fig:bioclipacc}
\end{figure}

Whether trained for robustness or not, accuracy curves such as those in Figure~\ref{fig:bioclipacc} form a logistic curve. Our hyperparameters for model robustness were chosen using a grid search similar to optimizing a model for test accuracy. Note that optimizing for test accuracy does not, in general, optimize for robustness to pruning. These specialized training regimes typically garner test accuracy for the unpruned model that is competitive with standard hyperparameter choices. Particularly important hyperparameters to tune for this purpose are batch size, the $\ell_2$-regularization constant, and training epochs.

\subsection{Hardware-Specific Speedup Analysis}
\label{sec:hardware-speedup}

We assess BioCLIP’s performance on three distinct compute environments to illustrate how pruning translates into inference speedups under varying resource constraints in the following settings: 


\paragraph{Raspberry Pi 4B} On the most constrained hardware, pruning yields the largest gains. Speedup reaches approximately $1.5\times$ at $0.3$ pruning ratio. The impact here is more pronounced due to the dominance of inference costs rather than overhead.

\paragraph{Ryzen 9 5950X} This high-end CPU benefits moderately, achieving up to $1.17\times$ speedup at $0.3$ pruning ratio. The diminished slope compared to the Pi is explained by the increased overhead and greater core count.

\paragraph{RTX 4070 GPU} While the GPU sees benefits from pruning, the overall gains (about $1.14\times$ at $0.3$ pruning ratio) are relatively modest. This shallower slope arises from fixed overheads and the architecture’s inherent parallelism: GPU kernels cannot fully utilize the newly available resources, limiting the device speedup. GPU threads unlike CPU must operate in lockstep, making it harder to utilize freed resources.
\linebreak

\noindent The results for the above hardware settings can be seen in Figure~\ref{fig:cross-platform-speedup}. Faster devices exhibit proportionally smaller performance gains per increment of pruning, due to hardware-specific overheads that remain constant regardless of model size. Nonetheless, all tested platforms maintain roughly linear, predictable scaling with pruning ratio. This reliability informs the design of our dynamic pruning controller, enabling it to balance accuracy and latency across heterogeneous devices.

\subsection{End-to-End Performance Under Real Workload}
\begin{figure}[htb]
\centering
\includegraphics[width=1\linewidth]{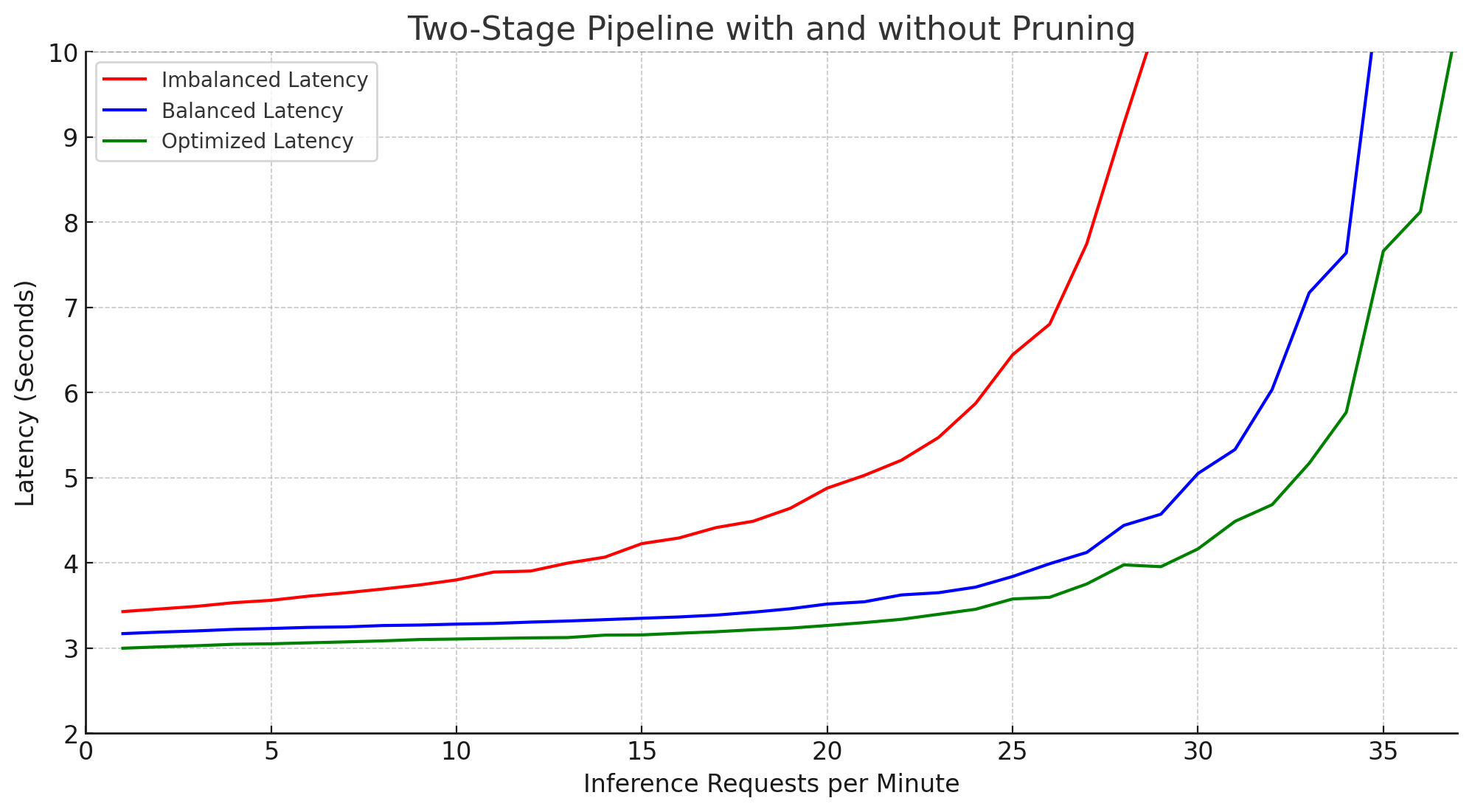}
\caption{BioCLIP pipeline with varying pruning levels and under many arrival rates}\label{fig:bioclipspeed}
\end{figure}

We use real camera-trap traces \cite{kline2024characterizing}, the BioCLIP model made to label the images 
\cite{stevens2024bioclip}, and edge hardware similar to the real camera-trap study deployment (Pi 4B) to evaluate our system. This experiment shows how our system would perform in a real animal ecology setting. 

First, our system generates benchmark curves, illustrated in Figures~\ref{fig:cross-platform-speedup} and \ref{fig:bioclipacc}.
Next, our system begins accepting inference requests, informed by real-world traces using the method described in \citep{kline2024characterizing}. Our camera setup generates data in intense bursts, so even though our average utilization may be low, it will experience transient spikes. Our clever placement leaves us with approximately a $14$\% load imbalance between our two nodes.  Under this realistic scenario, using pruning to balance, we approximately halve our latency while keeping accuracy above $80$\%.

\label{sec:discussion}

\section{Future Work}\label{sec:future}
A clear direction forward from this work is to showcase our training regimes and dynamic pruning system in common experimental settings such as ResNet or VGG models trained on standard datasets including CIFAR10. A rigorous demonstration of the principles of our system and training process will show that our ideas plausibly generalize to any visual model ran on the edge. In the future we hope to expand significantly on the pruning-aware training to show that higher pruning ratios than those of Figure~\ref{fig:bioclipacc} can be attained for widely available models.

Another future direction for our work is to integrate reinforcement learning into the dynamic pruning controller. In Section~\ref{sec:design}, we describe a reactive pruning controller that depends on hysteresis and current performance measurements. Reinforcement learning will allow our controller to act \emph{proactively} and thus anticipate when pruning may be required.
\section{Conclusion}
\label{sec:conclusion}


We have presented environment-aware dynamic pruning, a mechanism that enhances distributed edge inference through real-time structural adaptation. Our key insight is that neural network pruning, traditionally used as a static optimization technique, can serve as a powerful runtime load-balancing mechanism - both for rectifying inherent pipeline imbalances and handling transient device slowdowns. 
By carefully designing pruning-aware training procedures and lightweight monitoring systems, we enable immediate post-deployment adaptation with reduced accuracy loss. Our evaluation demonstrates significant benefits across heterogeneous edge deployments. In real-world experiments with camera trap workloads running on Raspberry Pi clusters, dynamic pruning improves end-to-end latency by up to $1.75\times$ while maintaining validation accuracy for the requisite pruning rates. The approach proves particularly valuable for resource-constrained devices, where even modest pruning ratios can substantially reduce computational bottlenecks.

Our system performs best when device utilization or pipeline imbalance are high. Due to our simple implementation of SLO violation detection and pruning, the overhead of our system is negligible compared to gains in total inference speed. Throughout this work, we have focused on making dynamic pruning practical for real-world deployments. While we evaluated on wildlife monitoring applications, the techniques we developed are likely to be applicable to many scenarios involving heterogeneous hardware, unpredictable workloads, or imperfect slice placements.



\bibliography{main}

\end{document}